\documentclass[twocolumn]{aastex701}
\usepackage{CJKutf8}

\shorttitle{Abundance--$T_{\text{c}}$ Correlations are Unrelated to
Planet Formation}
\shortauthors{Wang \& Schlaufman}

\received{July 18, 2026}
\revised{Oct 27, 2026}
\accepted{Nov 11, 2026}

\defcitealias{Abdurrouf2022}{Abdurro'uf et al. 2022}

\submitjournal{AAS Journals}

\graphicspath{{./}{figures/}}

\begin{document}
\begin{CJK*}{UTF8}{gbsn}

\title{Elemental Abundance Trends with Condensation Temperature are
Unrelated to Planet Formation}

\author[0000-0002-6379-3816]{Le-Chris Wang~(王乐)}
\affiliation{William H.\ Miller III Department of Physics \& Astronomy,
Johns Hopkins University, 3400 N Charles St, Baltimore, MD 21218, USA}
\affiliation{Department of Astrophysical Sciences, Princeton University,
4 Ivy Ln, Princeton, NJ 08544, USA}
\email[show]{lechris.wang@princeton.edu}

\author[0000-0001-5761-6779]{Kevin C.\ Schlaufman}
\affiliation{William H.\ Miller III Department of Physics \& Astronomy,
Johns Hopkins University, 3400 N Charles St, Baltimore, MD 21218, USA}
\email{kschlaufman@jhu.edu}

\begin{abstract}

\noindent
It has been suggested that the photosphere of the Sun is depleted in
the abundances of the refractory elements relative to the photospheres
of nearby solar twin stars and that the degree of depletion depends on
condensation temperature $T_{\text{c}}$.  While the formation of the
solar system's terrestrial and/or giant planets may play some role in the
refractory depletion of the solar photosphere, it has also been shown that
other processes can produce similar results.  If the terrestrial and/or
giant planet formation explanations are correct, then the depletion of
refractory elemental abundances in the photospheres of terrestrial/giant
exoplanet host stars should be more pronounced and therefore observable
even without achieving the highest possible elemental abundance inference
precisions in systems with host star surface convective zones less
massive than the Sun's.  For stars that host Kepler-discovered systems
of terrestrial exoplanets and have low-mass surface convective zones, we
do not observe the dependence of depletion on $T_{\text{c}}$ predicted by
the terrestrial planet formation scenario.  For a sample of stars that has
been uniformly searched with the Doppler technique for giant exoplanets,
we likewise show that giant exoplanet host stars are not depleted but
enhanced by about $0.0063\pm 0.0004$ dex in refractory abundances compared
to stars of similar masses and metallicities without giant exoplanets.
Since gaps opened in protoplanetary disks by giant planets are thought
to produce photospheric refractory depletion in their host stars, our
observations imply that at constant metallicity giant planets are more
likely to form in disks with high refractory abundances.

\end{abstract}

\keywords{\uat{Exoplanets}{498} ---
\uat{Extrasolar gaseous giant planets}{509} ---
\uat{Extrasolar rocky planets}{511} --- \uat{Planet formation}{1241} ---
\uat{Planet hosting stars}{1242} --- \uat{Stellar convection envelopes}{299}}

\section{Introduction}\label{sec:intro}

Careful differential elemental abundance inferences with precisions
approaching 0.01 dex indicate that refractory elements are
depleted by about 0.08 dex or 20\% relative to volatile elements
in the solar photosphere when compared to nearby solar twin
stars \cite[e.g.,][]{2009ApJ...704L..66M, 2009A&A...508L..17R}.
Those authors also found that the degree of depletion of individual
refractory elements increases with the temperature at which 50\%
of that element could be expected to condense \citep[referred to as
$T_{\text{c}}$;][]{2003ApJ...591.1220L}.  While this conclusion has been
been confirmed by several studies \citep[e.g.,][]{2010A&A...521A..33R,
2010ApJ...724...92C, 2010MNRAS.407..314G, 2011MNRAS.416L..80G,
2018ApJ...865...68B,Sun2025a,Sun2025b}, its accuracy has also been
questioned \citep[e.g.,][]{2010ApJ...720.1592G, 2013A&A...552A...6G,
2014A&A...564L..15A}.

\citet{2009ApJ...704L..66M} proposed that the solar photosphere's
low refractory-to-volatile ratio could be attributed to fact that
approximately two Earth masses of refractory elements initially present
in the protosolar nebula were sequestered in Mercury, Venus, Earth,
and Mars by the terrestrial planet formation process.  Because they
were locked up in terrestrial planets, those refractory elements were
prevented from accreting onto the young Sun and raising its photospheric
refractory-to-volatile ratio to the value observed in nearby solar twin
stars.  \citet{2009ApJ...704L..66M} and \citet{2009A&A...508L..17R} went
on to suggest that low refractory-to-volatile ratios in the photospheres
of solar twin stars could be interpreted as evidence for the existence
of terrestrial exoplanet systems, thereby circumventing the difficulty
of detecting terrestrial exoplanet systems using the Doppler, transit,
or direct imaging techniques.

While \citet{2009A&A...508L..17R} found that only about 15\% of
nearby solar twin stars display the abundance pattern they proposed
to be indicative of terrestrial planet formation, the frequency of
small-radius exoplanets in the Kepler was found by \citet{Fressin2013} to
be significantly higher with more than 26\% of stars hosting at least one
planet in the exoplanet radius and orbital period ranges $0.8~R_{\oplus}
\lesssim R_{\text{p}} \lesssim 2.0~R_{\oplus}$ and 0.8 days $\lesssim P
\lesssim 145$ days.  \citet{2021ApJ...907..116N} subsequently found that
at least 70\% of a sample of 1,700 solar-analog stars from the Apache
Point Galactic Evolution Experiment's second phase (APOGEE-2) displays
a refractory-to-volatile ratio similar to the solar photosphere though,
further complicating the interpretation of this abundance--$T_{\text{c}}$
pattern as a signature of terrestrial planet formation.

Other explanations unrelated to terrestrial planet formation for the
low refractory-to-volatile ratio in the solar photosphere compared
to nearby solar twin stars have been put forward.  The tidal
disruption and accretion of a super Earth-mass exoplanet that
occurs after a solar twin star reaches the zero-age main sequence
could boost that star's refractory-to-volatile ratio from the
low solar value to the value observed in most nearby solar twin
stars \citep[e.g.,][]{2011ApJ...740...76R, 2015A&A...582L...6S,
2021NatAs...5.1163S, 2018ApJ...854..138O}.  This explanation would
require the tidal disruption and accretion of super Earth-mass
exoplanets to be common though.  While \citet{2021NatAs...5.1163S} and
\citet{FLiu2024} advocated for a planet disruption and accretion rates of
20\%--35\% and at least 8\% respectively, \citet{2023MNRAS.521.2969B}
and \citet{Cheng2026} suggested a tidal disruption and accretion
rate of about 1\%.  \citet{2023MNRAS.521.2969B} emphasized that the
inhomogeneous abundance inferences and sample selection criteria used
by \citet{2021NatAs...5.1163S} could lead to the higher values.  The
observation by \citet{2024AJ....168..109S} that the ultra-short-period
super Earth-mass exoplanets most susceptible to tidal disruption
and accretion orbit the oldest stars disfavors the tidal disruption
and accretion scenario, as even young solar twin stars in the solar
neighborhood have super-solar refractory-to-volatile ratios.

Galactic chemical evolution has also been proposed as the best explanation
for the low refractory-to-volatile ratio in the solar photosphere
compared to nearby solar twin stars \citep[e.g.,][]{2013A&A...552A...6G,
2014A&A...564L..15A, 2015A&A...579A..52N, 2016A&A...585A.152S}.
\citet{2018ApJ...865...68B} sought to correct for Galactic chemical
evolution and recovered a low refractory-to-volatile ratio in the solar
photosphere, though \citet{2022MNRAS.512.3684C} emphasized that possible
refractory depletion of the solar photosphere depends sensitively
on the selection of comparison stars. Using the same solar-twin
sample and the \citet{2018ApJ...865...68B} age--abundance relations,
\citet{Babatsikos2026} instead compared the abundances of 18 elements
simultaneously with models for Galactic chemical evolution, intrinsic
abundance scatter, and planet engulfment.  They found that Galactic
chemical evolution explains the apparent chemical difference between
the Sun and most solar twins.

While \citet{2009ApJ...704L..66M} and \citet{2009A&A...508L..17R}
argued that the formation of the solar system's terrestrial planets
was the best explanation for the low refractory-to-volatile ratio
in the solar photosphere compared to nearby solar twin stars,
\citet{2020MNRAS.493.5079B} and \citet{2023A&A...676A..87H} instead
identified the formation of the giant planet Jupiter as the more important
factor.  Giant planets like Jupiter are capable of opening gaps in their
parent protoplanetary disks, and these gaps may then prevent solids
outside the orbits of newly formed giant planets from migrating inwards
and eventually accreting onto their host stars.  If gap-opening giant
planets form early in the evolution of their parent protoplanetary disks,
then they could trap more than $100~M_{\oplus}$ of solids outside their
orbits.  This sequestration of solids at large orbital separations could
easily create a low refractory-to-volatile ratio in the photospheres of
their host stars similar to the value observed in the solar photosphere.
In support of their argument, \citet{2020MNRAS.493.5079B} also noted
the similar occurrence of refractory depletion and long-period giant
exoplanets \citep[e.g.,][]{2008PASP..120..531C, 2021ApJS..255...14F}.
However, recent observational studies do not support the prediction
that giant-planet host stars should preferentially display solar-like
refractory depletion.  \citet{Carlos2025} found no relationship
between the presence of giant planets and the Sun's peculiar abundance
pattern, while \citet{Sun2025a, Sun2025b} found no clear dependence of
abundance--$T_{\text{c}}$ trends on planet type.

Both the terrestrial/giant planet formation scenarios predict that the
host stars of terrestrial/giant exoplanets should have photospheric
refractory-to-volatile ratios similar to the depleted solar value.
If the accretion that causes solar twin stars to have enhanced
refractory-to-volatile ratios occurs after they reach the zero-age
main sequence, then one prediction is that at constant sequestration
mass the amplitude of refractory-to-volatile depletion should be
inversely proportional to the mass of a star's surface convective
envelope.  \citet{2009A&A...508L..17R} suggested that for solar
twin stars with solar-like surface convective envelopes abundance
inference precisions of 0.03 dex or better are necessary to discern the
relationship between refractory elemental abundance and $T_{\text{c}}$.
The need for abundances precisions of 0.03 dex or better was confirmed
by \citet{2023AJ....165..178B} who showed that the abundances of
Kepler-discovered exoplanet host stars are indistinguishable from the
abundances of similar stars without Kepler-observed exoplanets, at
least down to the precision floor of their data at about 0.04 dex.
These results and the more sophisticated arguments presented in
\citet{2018A&A...618A.132K} emphasize the importance of focusing
investigations of the relationship between planet formation and host star
photospheric abundances on planet-hosting stars with low-mass surface
convective envelopes for which the relationship between refractory
elemental abundance and $T_{\text{c}}$ should be observable even with
lower precision elemental abundance inferences.

In this article, we use stellar evolution models to assemble a sample of
solar-type dwarf stars with low-mass surface convective envelopes and
previously published precise elemental abundance inferences that were
also observed by Kepler to host terrestrial exoplanet systems.  We next
use stellar evolution models to isolate a sample of solar neighborhood,
solar-type dwarf stars with low-mass surface convective envelopes and
previously published precise elemental abundance inferences that have
also been uniformly searched for giant exoplanets.  We then use both
data sets to evaluate the level of refractory-to-volatile depletion
in the samples, and we find that neither sample displays the depletion
predicted by the terrestrial/giant planet formation explanations for the
solar photosphere's low refractory-to-volatile ratio.  We describe our
stellar modeling in Section \ref{sec:Surfcz} and our sample selection
techniques in Section \ref{sec:sample} and our statistical evaluation of
the tests described above in Section \ref{sec:analysis}.  We discuss the
implications of our work in Section \ref{sec:discussion} and summarize
our conclusions in Section \ref{sec:conclusion}.

\section{Stellar Evolution Modeling}\label{sec:Surfcz}

We seek to evaluate the possible role of terrestrial and giant planet
formation on the solar photosphere's low refractory-to-volatile ratio
and its relationship between elemental abundance and condensation
temperature.  To identify stars that could be expected to evince the
smallest refractory-to-volatile ratios and the strongest elemental
abundance--$T_{\text{c}}$ relationships if terrestrial or giant
planet formation were responsible for those features, we calculate
the mass in the surface convective envelope of stars as a function
of the four independent variables of stellar evolution mass, age,
composition, and rotation using the state-of-the-art stellar
evolution code Modules for Experiments in Stellar Astrophysics
\citep[MESA;][]{2011ApJS..192....3P,2013ApJS..208....4P,
2015ApJS..220...15P, 2018ApJS..234...34P,2019ApJS..243...10P,
2023ApJS..265...15J}.

We adopt a MESA configuration that mimics the configuration
used in the generation of the MESA Isochrones and Stellar Tracks
\citep[MIST;][]{2016ApJ...823..102C, 2016ApJS..222....8D}.  Though
MIST provides the time evolution of many stellar properties over a
dense grid of mass and metallicity, it does not consistently provide
the masses of stars' surface convective envelopes $M_{\text{cz}}$.
The stellar evolution parameters adopted by MIST have been optimized
to reproduce the evolution of the Sun and stars in open clusters;
we therefore adapt the MIST configuration used with MESA v7503
to MESA revision 22.11.1 to the extent possible.  We start with
\citet{2009ARA&A..47..481A} solar abundances.  We utilize opacities from
OPAL \citep{1996ApJ...464..943I} and AESOPUS \citep{2009A&A...508.1539M}
plus the equations of state from OPAL \citep{2002ApJ...576.1064R},
SCvH \citep{1995ApJS...99..713S}, and HELM \citep{2000ApJS..126..501T}
as provided by default in MESA.  Because we are only interested in
main sequence stellar evolution, we accelerate our calculations by
using the \texttt{basic.net} nuclear reaction network instead of the
more comprehensive \texttt{mesa\_49.net} network used in MIST.  We use
the Ledoux criterion to identify convective regions that we then model
with the mixing length theory of convection with mixing length parameter
$\alpha_{\text{MLT}} = 1.82$ calibrated to the Sun.  We account for atomic
diffusion using the framework proposed by \citet{1994ApJ...421..828T}.
We treat convective overshoot with an exponential-decay model parameter
0.0172 from a convective core to a radiative envelope and 0.016 from
a convective envelope to a radiative core; the former is calibrated to
the open cluster M 67 and the latter is calibrated to the Sun.  We use
an outer boundary condition based on \texttt{ATLAS12} model atmospheres
\citep{1970SAOSR.309.....K,1993sssp.book.....K}.  We neglect mass loss
entirely as it is unimportant during the evolution of solar-type stars
from protostars to the terminal-age main sequence (TAMS).

As is common practice, we account for stellar rotation
only after a star reaches the zero-age main sequence (ZAMS)
\citep[e.g.,][]{2008Ap&SS.316...43E, 2016ApJ...823..102C}.  MIST only
accounts for the rotation of stars with initial masses $M_{\ast}
> 1.2$ $M_{\odot}$ and increases rotation from zero to its maximal
value over the mass interval $1.2~M_{\odot} < M_{\ast} < 1.8~M_{\odot}$
to compensate for the neglect of magnetic braking mechanisms in MESA.
We likewise ignore the small and likely insignificant impact of magnetic
braking on the internal structures of solar-type stars during their main
sequence evolution.

Our MESA calculations quantify the evolution of solar-type stars from
the protostar stage to the TAMS over a range of mass, metallicity, and
rotation period.  We calculate the evolution of stars with $M_{\ast}
\in \{0.6,0.8,1.0.1.2\}~M_{\odot}$, metal mass fractions $Z_{\ast}$ with
$\log_{10}(Z_{\ast}/Z_{\odot})$ from $-1$ to $+0.5$ in increments of 0.1,
and ZAMS rotation periods $P_{\text{rot}} \in \{1,10,\infty\}$ days.
Using these evolution calculations, we record surface convective
zone masses $M_{\text{cz}}$ from age $\log_{10}(\tau/\text{yr})
= -4$ to $\log_{10}(\tau/\text{yr}) = 11$ or the TAMS in steps of
$\log_{10}(\tau/\text{yr}) = 0.5$.  Since MESA reports the evolution of
stellar structure at irregular time intervals, we interpolate quantities
of interest with cubic splines to evaluate those quantities at any time.

\begin{figure*}
\plotone{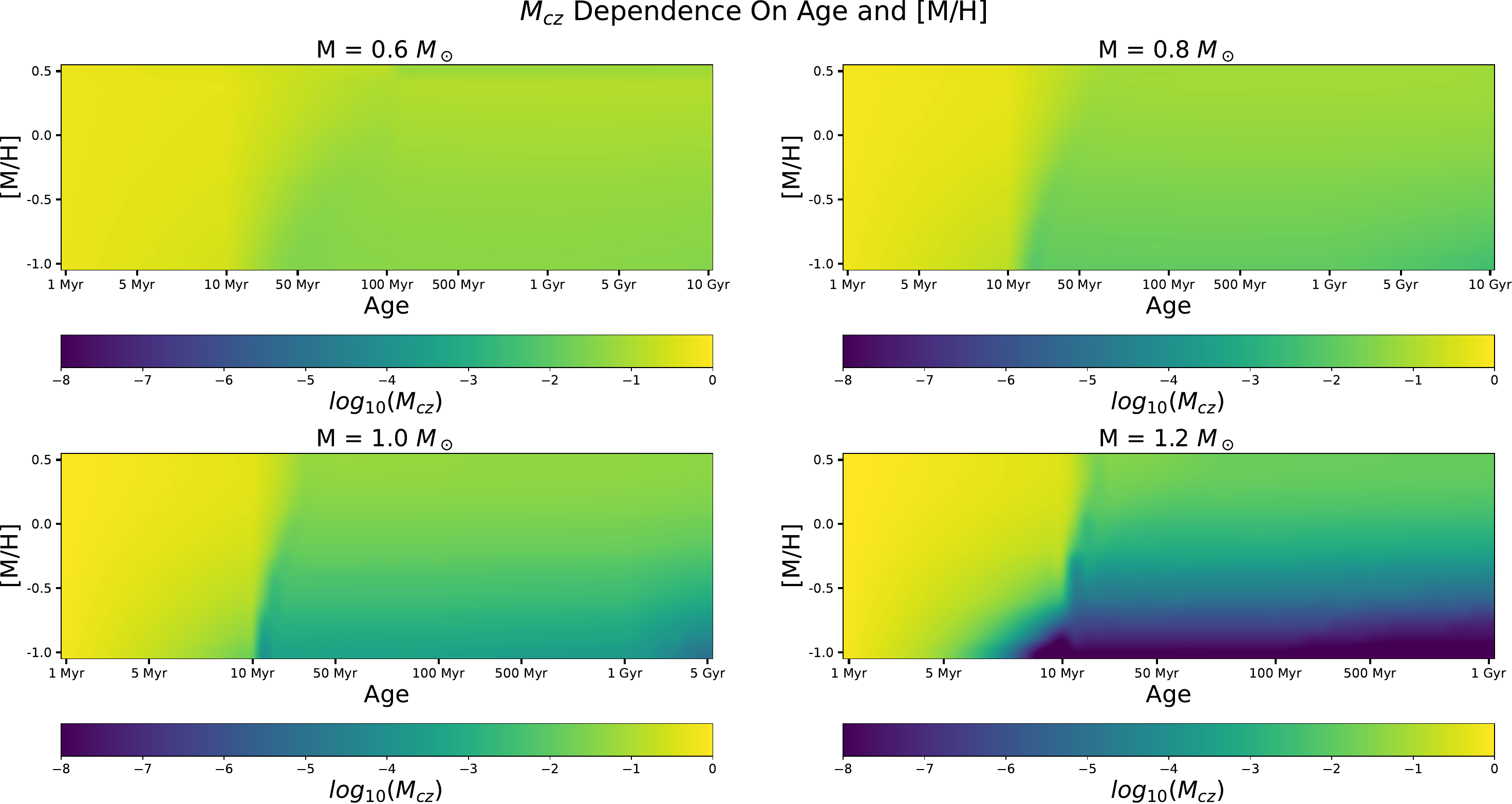}
\caption{Heat maps illustrating the outcomes of our MESA calculations
of the evolution of surface convective zone masses $M_{\text{cz}}$
with age $\tau$ as a function of metallicity $[\text{M/H}]$ for four
stellar masses $M_{\ast} \in \{0.6,0.8,1.0,1.2\}~M_{\odot}$ all with
zero-age main sequence (ZAMS) rotation periods $P_{\text{rot}} = 1$ day.
$M_{\text{cz}}$ starts to shrink at $\tau \approx 10$ Myr as stars
approach the ZAMS.  At all ages, $M_{\text{cz}}$ is a decreasing function
of mass and an increasing function of metallicity.\label{fig:heatmap}}
\end{figure*}

\begin{figure*}
\plottwo{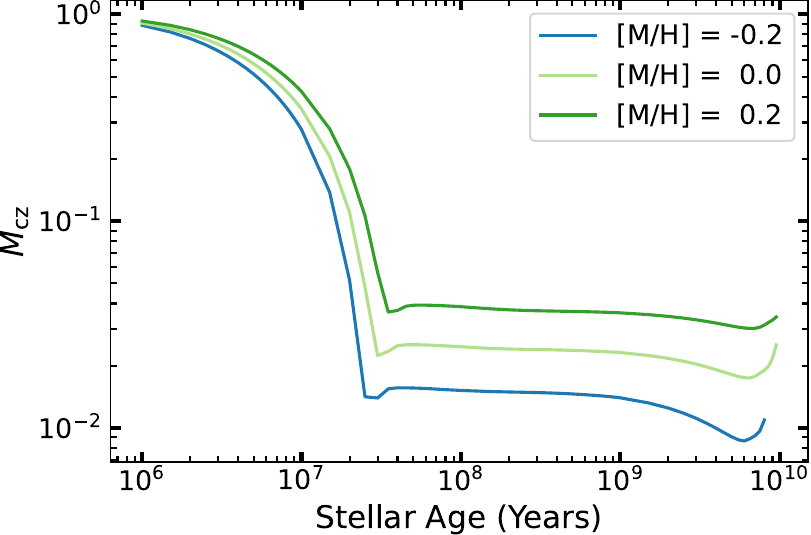}{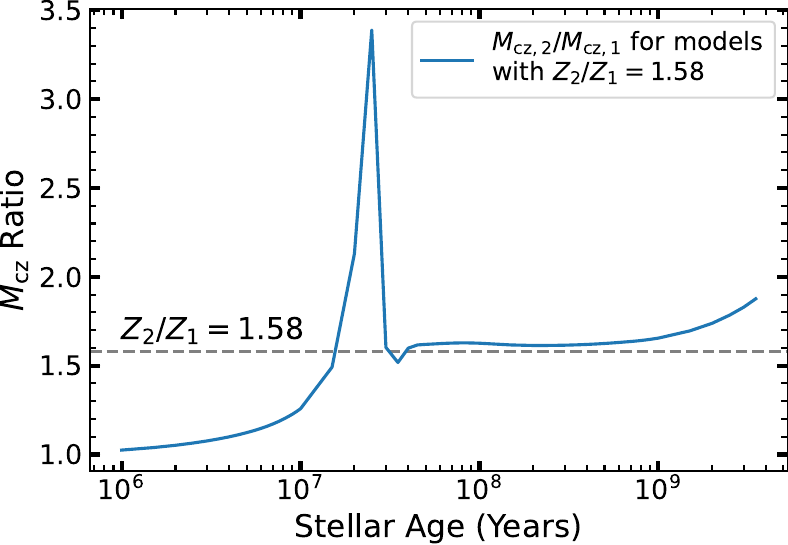}
\caption{Time evolution of $M_{\text{cz}}$ predicted by our MESA
calculations for solar-mass stars as a function of metallicity with
rotation period $P_{\text{rot}} = 10$ day at the ZAMS.  Left: effect of
metallicity on $M_{\text{cz}}$ evolution for stars with $[\text{M/H}]\in
\{-0.2, 0.0, 0.2\}$.  Right: $M_{\text{cz}}$ ratio for models with
$([\text{M/H}]_{2} = 0.0)/([\text{M/H}]_{1} = -0.2) \Rightarrow
Z_{2}/Z_{1} = 1.58$ showing that $M_{\text{cz}}$ scales roughly linearly
with metallicity for solar-type stars.\label{fig:evolution}}
\end{figure*}

We show in Figure \ref{fig:heatmap} that $M_{\text{cz}}$ is an increasing
function of metallicity and decreasing function of mass at all times
for solar-type stars.  In the mass and metallicity ranges $0.6~M_{\odot}
\lesssim M_{\ast} \lesssim 1.2~M_{\odot}$ and $-1.0 \lesssim [\text{Fe/H}]
\lesssim +0.5$ relevant for the host of exoplanets detected by Kepler
or ground-based Doppler surveys, stars younger than a few Myr are fully
convective.  As stars approach the ZAMS, the masses of their surface
convective zones decrease over about 10 Myr, with the fastest evolution
for the most massive and lowest metallicity stars.  They then asymptote
to mass- and metallicity-dependent values that remains roughly constant
through stars' main sequence evolution.  These trends are a consequence of
higher opacities in the envelopes of lower mass or more metal-rich stars
that produce steeper radiative temperature gradients and therefore larger
surface convective zones.  In Figure \ref{fig:evolution} we focus on the
effect of metallicity on $M_{\text{cz}}$ and show that for solar-type
stars in the mass and metallicity ranges $0.6~M_{\odot} \lesssim M_{\ast}
\lesssim 1.2~M_{\odot}$ and $-1.0 \lesssim [\text{Fe/H}] \lesssim +0.5$
$M_{\text{cz}}$ scales roughly linearly with metallicity.

\begin{figure*}
\plotone{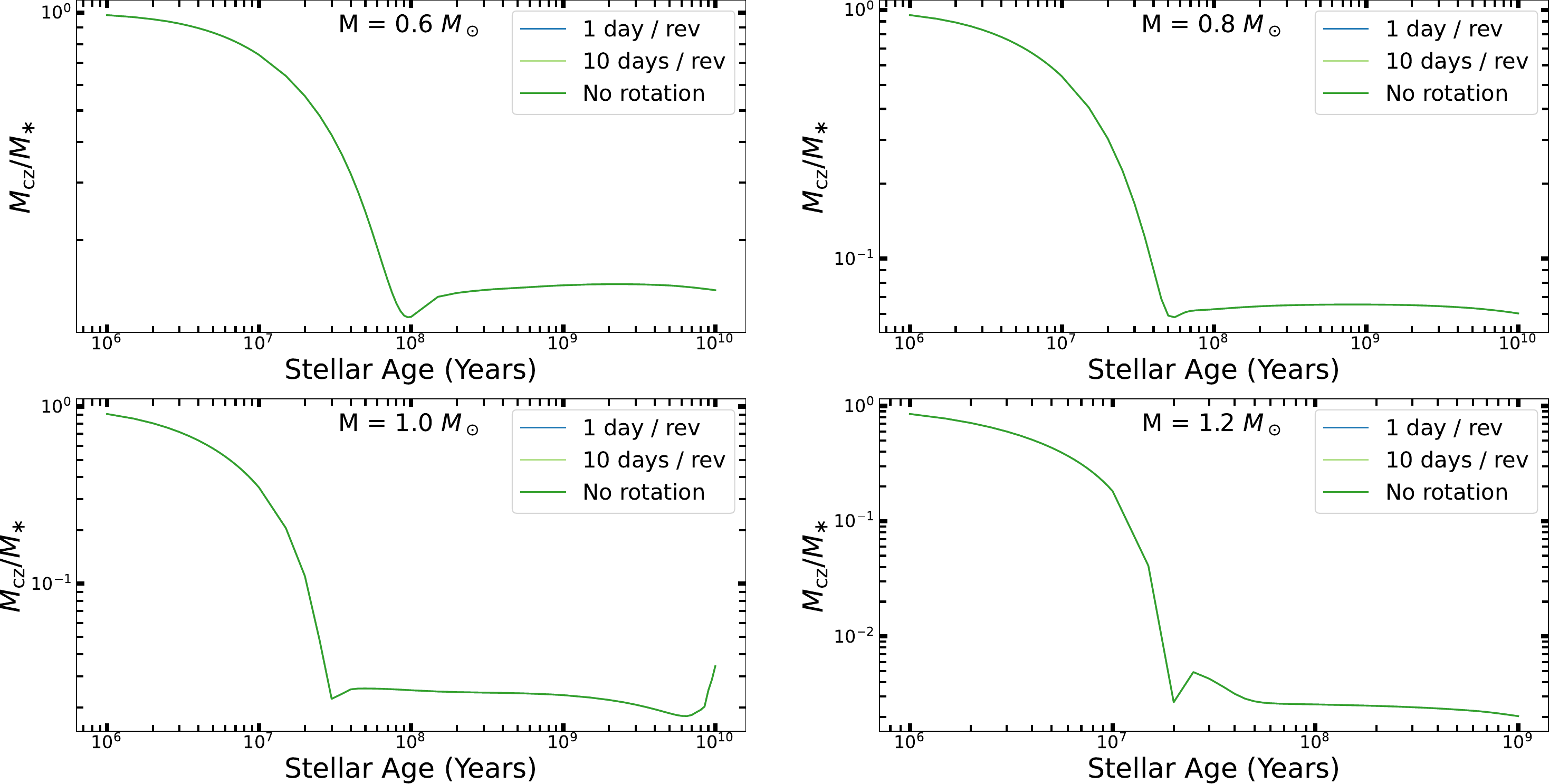}
\caption{Surface convective zone mass normalized to total stellar
mass $M_{\text{cz}}/M_{\ast}$ predicted by our MESA simulations for
solar composition stars as a function of stellar mass $M_{\ast} \in
\{0.6,0.8,1.0,1.2\}~M_{\odot}$ and rotation period $P_{\text{rot}}
\in \{1,10,\infty\}$ days at the ZAMS.  In the plausible ZAMS rotation
period range $1~\text{day} \lesssim P_{\text{rot}} \lesssim \infty$
days, rotation does not affect $M_{\text{cz}}$.\label{fig:rotation}}
\end{figure*}

In the mass and metallicity ranges $0.6~M_{\odot} \lesssim M_{\ast}
\lesssim 1.2~M_{\odot}$ and $-1.0 \lesssim [\text{Fe/H}] \lesssim +0.5$
$M_{\text{cz}}$, stellar rotation rate barely impacts $M_{\text{cz}}$
(Figure \ref{fig:rotation}).  Even though rotation provides additional
support against convective instability, it is only relevant for regions
with Prandtl number $\text{Pr} \geq 1$.  Stellar convective zones have
$\text{Pr} \ll 1$ \citep[e.g.,][]{2022ApJS..262...19J}, so rotation is
inconsequential for the calculation of solar-type stars' $M_{\text{cz}}$.
Likewise, rotationally induced mixing is unimportant for the evolution
of the structure of solar-type stars.  The approximation of spherical
symmetry is appropriate for solar-type stars across the rotation period
values we study that are typical of the rotation periods of solar-type
stars, as even the shortest periods we study imply low rotational energies
compared to the gravitational potential energies.

\section{Sample Selection}\label{sec:sample}

Giant planets and the oligarchs that participate in the giant impact stage
of terrestrial planet formation form while their parent protoplanetary
disks are present.  Most protoplanetary disks dissipate in just a few
million years \citep[e.g.,][]{2001ApJ...553L.153H}.  Solar-type stars
with masses metallicities in the ranges $0.6~M_{\odot} \lesssim M_{\ast}
\lesssim 1.2~M_{\odot}$ and $-1.0 \lesssim [\text{Fe/H}] \lesssim
+0.5$ relevant for the host stars of exoplanets detected by Kepler or
ground-based Doppler surveys do not reach the ZAMS and therefore remain
at least substantially convective while their protoplanetary disks are
present (Figure \ref{fig:heatmap}).  The impacts of sequestration or
accretion of refractory elements while a protoplanetary disk is present
are consequently diluted by a large fraction of the mass of the star
hosting the protoplanetary disk.

Assuming the \citet{2009ARA&A..47..481A} protosolar mass fraction $Z =
0.0142$ of which only about 23\% is in refractory elements, sequestering
$10~M_{\oplus}$ of refractory material in the progenitors of terrestrial
planets could decrease the refractory abundances of the solar photosphere
by an unobservably small 0.004 dex.  On the other hand, sequestering
$100~M_{\oplus}$ of refractory material in the outer protosolar nebula
as a consequence of giant planet formation could decrease the refractory
abundances of the solar photosphere by a possibly observable 0.04 dex.
The implication is that while the sequestration of refractory elements in
the oligarchs that participate in the giant impact stage of terrestrial
planet formation is unlikely to be responsible for the solar photosphere's
low refractory-to-volatile ratio, the formation of Jupiter and its
possible sequestration of refractory elements in the outer part of
protosolar nebula cannot be ruled out by Figure \ref{fig:heatmap} alone.

The giant impact phase of terrestrial planet formation continued for
tens of Myr in the solar system \citep[e.g.,][]{2009GeCoA..73.5150K},
comparable to the time required for the Sun to reach the ZAMS and
realize the convective envelope mass it has had for its subsequent
main sequence evolution.  While the giant impact phase of terrestrial
planet formation is expected to produce a terrestrial planet system
with non-zero orbital eccentricities, the low orbital eccentricities of
the solar system's terrestrial planets suggest that their orbits were
circularized as a result of interactions with residual planetesimals.
These interactions would have stirred up the orbital eccentricities of
those residual planetesimals, increased their rate of collisions, and
ground them down to micron-size dust grains that could be blown out of
the solar system by radiation pressure.

If on the other hand terrestrial planet formation had not taken place,
it would be reasonable to expect a more massive disk of planetesimals
that experiences weaker collisional grinding.  In that scenario,
the inward drift of a few Earth masses of cm-size pebbles due to
Poynting–Robertson drag \citep{1903MNRAS..64A...1P, 1937MNRAS..97..423R}
or retrograde rotating planetesimals due to the Yarkovsky effect
\citep{1951PRIA...54..165O} could enhance the refractory-to-volatile
ratios of the photospheres of solar twin stars after they reach the ZAMS.
This would take hundreds of Myr and occur after solar twin stars realized
the convective envelope mass they have had for their subsequent main
sequence evolution.  If in this way terrestrial planet formation is
responsible for the solar photosphere's low refractory-to-volatile ratio
and its elemental abundance--$T_{\text{c}}$ relationship, then stars
with low-mass surface convective zones should show the strongest signals.

To investigate the scenarios outlined above and evaluate
the potential connection between terrestrial planet formation
and solar photosphere's low refractory-to-volatile ratio and its
elemental abundance--$T_{\text{c}}$ relationship, we need a sample of
terrestrial exoplanet host stars with accurate and precise fundamental
and photospheric stellar parameters as well as elemental abundance
inferences for many volatile and refractory elements.  To obtain
such a sample of stars, we join the \citet{2016ApJS..225...32B} and
\citet{2018ApJS..237...38B} catalogs of stellar parameters and elemental
abundance inferred from Keck/HIRES spectra \citep{1994SPIE.2198..362V}
using Spectroscopy Made Easy \citep[\texttt{SME};][]{2017A&A...597A..16P}
with the Kepler Data Release (DR) 25 Kepler Object of Interest (KOI) list
presented in \citet{2018ApJS..235...38T} hosted by the NASA Exoplanet
Archive \citep{2013PASP..125..989A,2025PSJ.....6..186C} as the table
\texttt{q1\_q17\_dr25\_koi}.  We first use Simbad to identify the Gaia
DR3 and Kepler Input Catalog (KIC) identifiers for all of the stars
in the \citet{2016ApJS..225...32B} and \citet{2018ApJS..237...38B}
catalogs.  For those stars not immediately resolvable by Simbad, we
manually identify the Simbad entry that corresponds to star listed in the
\citet{2016ApJS..225...32B} and \citet{2018ApJS..237...38B} catalogs.
Similarly, we manually retrieve Gaia DR3 source identifiers for those
stars that lack those data in Simbad.  We next remove duplicate entries
and merge the \citet{2016ApJS..225...32B} and \citet{2018ApJS..237...38B}
catalogs using Gaia DR3 source identifiers as the primary key for
each table.

We then merge the combined
\citet{2016ApJS..225...32B}/\citet{2018ApJS..237...38B} catalog with
the Kepler DR25 KOI list using KIC identifier.  After removing KOIs
dispositioned as false positives, we next use Kepler DR25 KOI list
planet-to-star radius ratios and the stellar radii in the combined
\citet{2016ApJS..225...32B}/\citet{2018ApJS..237...38B} catalog
to calculate updated planet radii.  We then identify stars that
host terrestrial exoplanet systems, or in other words, more than
one terrestrial exoplanet with $R_{\text{p}} < 1.7~R_{\oplus}$ and
$P < 16$ d.  We use the latter criterion because we seek a complete
terrestrial exoplanet sample and Kepler is only complete to the smallest
terrestrial exoplanets with $R_{\text{p}} \approx 0.5~R_{\oplus}$ out to
orbital periods $P \lesssim 16$ d \citep[e.g.,][]{2019AJ....158..109H}.
We next use the \citet{2019PNAS..116.9723Z} Earth-composition radius--mass
relation to calculate the total mass in observed terrestrial exoplanets
in each system.\footnote{\citet{2025ApJ...993..235R} have recently
demonstrated that most terrestrial exoplanets have Earth-like internal
structures, so our assumption of Earth-like compositions is a good one.}
This procedure results in a sample of 72 terrestrial exoplanet systems
with host star photospheric elemental abundances for carbon, nitrogen,
oxygen, sodium, magnesium, aluminum, silicon, calcium, titanium, vanadium,
chromium, manganese, iron, nickel, and yttrium with uncertainties in
the range $0.02 < [\text{X/H}]_{\text{unc}} < 0.08$.

As we will describe in Section \ref{sec:analysis}, our analysis approach
requires a sample of stars representative of the overall population that
produced our sample of terrestrial exoplanet host stars.  To meet that
need, we use a similar procedure to match via KIC identifiers the combined
\citet{2016ApJS..225...32B}/\citet{2018ApJS..237...38B} catalog with the
sample of stars in the Kepler field with Kepler light curves that were
searched for transiting exoplanets recorded in the NASA Exoplanet Archive
\texttt{keplerstellar} table.  This procedure results in a sample of 1399
Kepler-observed stars with the same precise stellar photospheric elemental
abundance inferences as the terrestrial exoplanet system host sample.

To ensure that our analyses are not subject to any elemental
abundance systematic uncertainties specific to the combined
\citet{2016ApJS..225...32B}/\citet{2018ApJS..237...38B} catalog,
we use a similar procedure to define both terrestrial exoplanet
host star and Kepler-observed star samples using data collected
during the third and fourth phases of the Sloan Digital Sky
Survey \citep[SDSS;][]{2011AJ....142...72E,2017AJ....154...28B}
as part of its Apache Point Observatory Galactic Evolution
Experiment \citep[APOGEE;][]{2017AJ....154...94M} and released
to the community as part of its DR17 \citep{2022ApJS..259...35A}.
The spectra on which the SDSS DR17 stellar parameters are based were
collected with the APOGEE spectrograph \citep{2013AJ....146...81Z,
2017AJ....154..198Z, 2019PASP..131e5001W, 2021AJ....162..302B} on the
New Mexico State University 1-m Telescope \citep{2010AdAst2010E..46H}
or the Sloan Foundation 2.5-m Telescope \citep{2006AJ....131.2332G}.
As part of SDSS DR17, these spectra were reduced and analyzed
with the APOGEE Stellar Parameter and Chemical Abundance Pipeline
\citep[ASPCAP;][]{2006ApJ...636..804A, 2015AJ....150..148H,
2015AJ....150..173N, 2016AJ....151..144G} using an $H$-band line
list, MARCS model atmospheres, and model-fitting tools optimized for
the APOGEE effort \citep{1998A&A...330.1109A, 2008A&A...486..951G,
2011ascl.soft09022H, 2012ascl.soft05004P, 2013ApJ...765...16S,
2021AJ....161..254S, 2015ApJ...798L..41C, 2015ApJS..221...24S,
2020AJ....160..120J}.  These SDSS DR17 APOGEE data all have Gaia DR3
source information, and we use our own table connecting Gaia DR3 source
identifier and KIC identifier to find KIC identifiers for each star.
We obtain stellar mass and radius information for this Kepler APOGEE
sample from \citet{2022AJ....163..128W} and then proceed as described
above.  This procedure results in a sample of 16 terrestrial exoplanet
system host stars and 2543 Kepler-observed stars with precise elemental
abundance inferences.  Both the terrestrial exoplanet system host star
and Kepler-observed star samples have photospheric elemental abundances
for carbon, oxygen, magnesium, aluminum, silicon, potassium, calcium,
manganese, iron, and nickel with uncertainties in the range $0.01 <
[\text{X/H}]_{\text{unc}} < 0.09$.

To test whether gap-opening giant planets are connected to low stellar
refractory-to-volatile ratios, we require two homogeneously defined
samples: stars hosting planets capable of opening gaps in their
protoplanetary disks and otherwise similar single stars without such
planets or stellar companions.  Both samples must also have accurate
stellar parameters and precise abundances for many volatile and
refractory elements.  To obtain those samples, we use the combined
\citet{2016ApJS..225...32B}/\citet{2018ApJS..237...38B} catalog with
planetary and stellar companions reported in \citet{2021ApJS..255....8R}
based on a homogeneous analysis of a large radial velocity archive.
We use the same procedure described above to obtain Gaia DR3 source
identifier information for the \citet{2021ApJS..255....8R} sample and then
merge the two data sets using Gaia DR3 source identifier as the primary
key for each table.  To identify exoplanets massive enough to open gaps
in their parent protoplanetary disks, we use the criterion given as
Equation (15) in \citet{2006Icar..181..587C} assuming protoplanetary
disk aspect ratios $h/a = 0.05$, Shakura-Sunyaev viscosity parameter
$\alpha = 10^{-3}$, and planet-to-star mass ratios $q = M_{\text{p}}
\sin{i} / M_{\ast}$\footnote{Even though Doppler measurements only
provide minimum masses, any exoplanet with a minimum mass sufficient
for gap opening would also be able to open a gap if its true mass is
more than its Doppler-inferred minimum mass}.  This procedure results
in a sample of 60 hosts of exoplanets with $M_{\text{p}} \sin{i} \leq
10~M_{\text{Jup}}$ massive enough to have opened a gap in their parent
protoplanetary disk and a sample of 224 single stars uniformly searched
for but lacking exoplanets.

\section{Analysis}\label{sec:analysis}

The two planet formation scenarios we study make different predictions
for host star elemental abundances, so we use different metrics
to evaluate them.  As explained in Section \ref{sec:tpsystem},
\citet{2010ApJ...724...92C} provided an explicit prescription relating
the mass of refractory material sequestered during terrestrial
planet formation to the expected abundance--$T_{\text{c}}$ slope.
It therefore enables a direct comparison between predicted and
observed slopes.  In contrast, while \citet{2020MNRAS.493.5079B} and
\citet{2023A&A...676A..87H} suggested that gap-opening giant planets would
suppress the accretion of refractory-rich solids and thereby produce
a relative depletion of refractory elements, they did not predict an
element-by-element abundance--$T_{\text{c}}$ relation.  We therefore test
the former using abundance--$T_{\text{c}}$ slopes and the latter using
the refractory-to-volatile abundance difference for giant planet-hosting
stars relative to matched analog stars that do not host giant planets.

\subsection{Terrestrial Exoplanet System Formation}\label{sec:tpsystem}

To evaluate the possible impact of the formation terrestrial
exoplanet systems on the elemental abundances of their host stars,
we use the samples of Kepler-discovered terrestrial exoplanet
systems and Kepler-observed stars with elemental abundances from the
combined \citet{2016ApJS..225...32B}/\citet{2018ApJS..237...38B}
and APOGEE DR17 catalogs described in Section \ref{sec:sample}.
\citet{2009A&A...508L..17R} quantified the degree of refractory element
depletion of a star relative to the Sun using the root mean square
slope $S_{\text{obs}}$ of refractory element abundance $[\text{X/H}]$
versus condensation temperature $T_{\text{c}}$.  According to this
definition, stars with solar composition have $S_{\text{obs}} = 0$, star
enriched relative to solar in refractory elements have $S_{\text{obs}}
> 0$, and stars depleted relative to solar in refractory elements
have $S_{\text{obs}} < 0$.  For stars in the solar neighborhood,
$S_{\text{obs}} \sim 10^{-3}$ dex K$^{-1}$.

To relate the observable $S_{\text{obs}}$ to fundamental stellar
parameters like mass, age, and composition, \citet{2010ApJ...724...92C}
assumed that any solar-like depletion of refractory elements is restricted
to the surface convective zone of a star and therefore that
\begin{eqnarray}
S_{\text{pred}} & = & S_{\text{max}} - \frac{\log_{10}[(M_{\text{cz}} + M_{\text{frac}})/M_{\text{cz}}]}{1300~\text{K} - 900~\text{K}}.\label{eqn:S}
\end{eqnarray}
In Equation (\ref{eqn:S}) $S_{\text{max}}$ is the slope observed
in a star that accretes no fractionated material assumed by
\citet{2010ApJ...724...92C} to have the value $S_{\text{max}} =
0.2$ to match the most refractory-enriched solar analogs reported
in \citet{2009A&A...508L..17R} while $M_{\text{frac}}$ is the
mass of refractory-fractionated materials accreted onto the star.
$S_{\text{max}}$ depends on both stellar sample and analysis approach,
and we describe later in this subsection how we calculate the value
of $S_{\text{max}}$ for our samples.  The quantity $M_{\text{frac}}$
is related to the total mass of terrestrial planet forming material
missing from a star
\begin{eqnarray}
M_{\text{rock}} & = & 0.0044~M_{\text{frac}}~10^{[\text{Fe/H}]},\label{eqn:rock}
\end{eqnarray}
roughly corresponding to the total mass of any terrestrial planet system
and/or asteroid belts in a system.  The factor 0.0044 in Equation
(\ref{eqn:rock}) is the mass fraction of rock-forming elements in
the solar photosphere as derived by \citet{2010ApJ...724...92C}, and
we assume in our analyses a similar relationship that scales only with
overall metallicity $[\text{Fe/H}]$.  In this \citet{2010ApJ...724...92C}
formalism, one can first infer the values of $S_{\text{obs}}$ for
terrestrial exoplanet systems using their host stars' photospheric
elemental abundances, then compare the values of $S_{\text{obs}}$ with the
values of $S_{\text{pred}}$ implied by the scenario where the formation
of those terrestrial exoplanet systems sequestered the masses observed
in terrestrial exoplanets that would have otherwise have accreted onto
their host stars.

\begin{figure*}
\plottwo{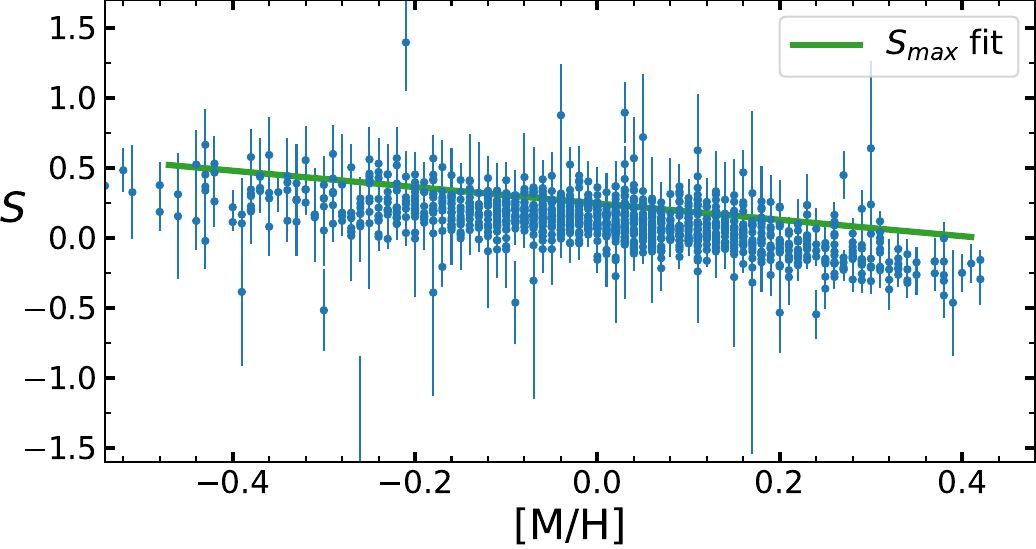}{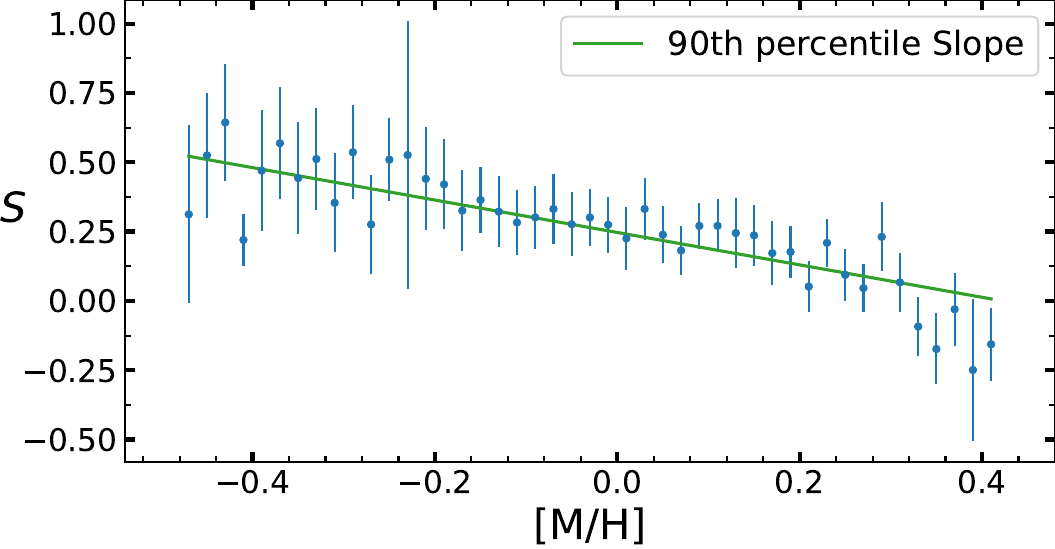}
\caption{Distribution of the \citet{2010ApJ...724...92C} $S_{\text{obs}}$
parameter that quantifies the degree of refractory depletion in a
stellar photosphere.  Left: the distribution of the $S_{\text{obs}}$
parameter for Kepler-observed stars with elemental abundance inferences
from the combined \citet{2016ApJS..225...32B}/\citet{2018ApJS..237...38B}
catalog shows that $[\text{M/H}]$ and $S_{\text{obs}}$ are anticorrelated.
Right: to identify a reasonable maximum value of the $S_{\text{obs}}$
parameter $S_{\text{max}}$ as a function of $[\text{M/H}]$, we bin in
0.02 dex bins the $[\text{M/H}]$ data plotted in the left-hand panel.
We calculate the 90th percentile of the $S_{\text{obs}}$ distribution
$S_{90}$ in each bin and then fit a line to the bin centers and the
$S_{90}$ values.  We use the resulting best-fit line to calculate
$S_{\text{max}}$ as a function of $[\text{M/H}]$.\label{fig:s_max}}
\end{figure*}

To obtain $S_{\text{max}}$ for our sample of Kepler-observed terrestrial
exoplanet systems, we first calculate the value of $S_{\text{obs}}$
for each individual star in our Kepler-observed samples based
on their inferred photospheric abundances.  As we show in Figure
\ref{fig:s_max}, as a consequence of Galactic chemical evolution
$S_{\text{obs}}$ is clearly a function of metallicity.  To calculate
the $S_{\text{max}}$ value appropriate for each metallicity, we bin
individual $S_{\text{obs}}$ values in bins 0.02 dex wide, calculate the
90th percentile of the $S_{\text{obs}}$ distribution $S_{90}$ in each
bin, and then fit using weighted least squares the line connecting
bin centers to the values of $S_{90}$ in each bin.  We illustrate
this calculation in Figure \ref{fig:s_max}.  We execute these steps
using the elemental abundances for Kepler-observed stars from both the
combined \citet{2016ApJS..225...32B}/\citet{2018ApJS..237...38B} and
APOGEE DR17 catalogs and validate our approach by confirming the values
$S_{\text{max}} = 0.24$ and $S_{\text{max}} = 0.26$ we infer in this way
at $[\text{M/H}] = 0.0$ are very similar to the value $S_{\text{max}}
= 0.2$ inferred by \citet{2010ApJ...724...92C} for the solar analogs at
$[\text{M/H}] = 0.0$.

To calculate $S_{\text{pred}}$ for each of the terrestrial exoplanet
system host stars in our samples, we follow the stellar evolution
calculation procedure described in Section \ref{sec:Surfcz}.  We first
use the stellar masses, metallicities, and ages reported in the combined
\citet{2016ApJS..225...32B}/\citet{2018ApJS..237...38B} or APOGEE
DR17 catalogs as inputs to rotationless MESA stellar evolution models
to calculate the current mass in the surface convective zone of each
terrestrial exoplanet system host star $M_{\text{cz}}$ as required by the
\citet{2010ApJ...724...92C} formalism.  We calculate $M_{\text{frac}}$
using Equation (\ref{eqn:rock}) with stellar metallicities from the
combined \citet{2016ApJS..225...32B}/\citet{2018ApJS..237...38B} or
APOGEE DR17 catalogs and total masses in terrestrial exoplanet systems
calculated using the \citet{2019PNAS..116.9723Z} Earth-composition
radius--mass relation as described in Section \ref{sec:sample}.  We then
use these values along with the value of $S_{\text{max}}$ described above
to calculate $S_{\text{pred}}$ for each terrestrial exoplanet system host
star under the assumption that the refractory material locked up in its
observed terrestrial exoplanet system would otherwise have accreted onto
the host star.  We note that the total masses in terrestrial exoplanets
we calculate are only lower limits on the total rocky masses potential
missing from host stars' surface convective zones, as there could be
terrestrial exoplanets unobserved by Kepler orbiting the host stars in our
samples.  The $S_{\text{pred}}$ values we infer in this way are therefore
upper limits on the true value of $S_{\text{pred}}$ for each system.

\begin{figure*}
\plotone{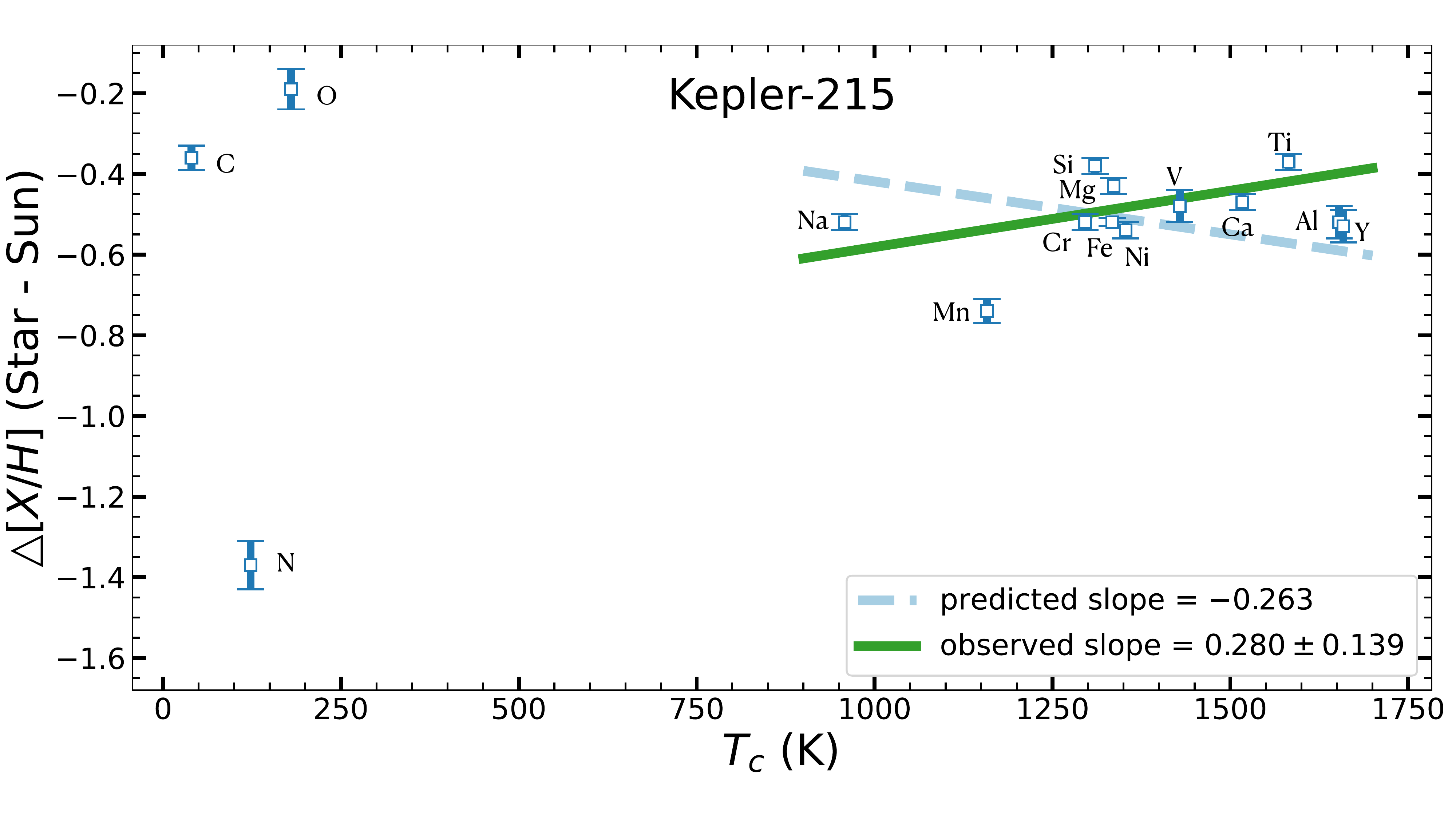}
\caption{Photospheric elemental abundances from the combined
\citet{2016ApJS..225...32B}/\citet{2018ApJS..237...38B} catalog as a
function of condensation temperature $T_{\text{c}}$ for the terrestrial
exoplanet system host star Kepler-215 that is representative of the 11
Kepler-discovered terrestrial exoplanet systems that we predict to show
the strongest refractory depletion signals.  We plot as the dashed blue
line the prediction from the \citet{2010ApJ...724...92C} formalism under
the assumption that the refractory material locked up in the observed
terrestrial exoplanets would otherwise have fallen onto the host star
and increased its photospheric refractory abundances.  For the refractory
elements with $T_{\text{c}} > 900$ K, we fit using weighted least squares
a linear relationship between $T_{\text{c}}$ and photospheric elemental
abundance and plot the result as the solid green line.  The slopes of the
two lines are very different, suggesting that terrestrial planet formation
in the Kepler-215 system did not result in the depletion of refractory
elements in the photosphere of its host star.\label{fig:Kepler}}
\end{figure*}

\begin{figure*}
\plotone{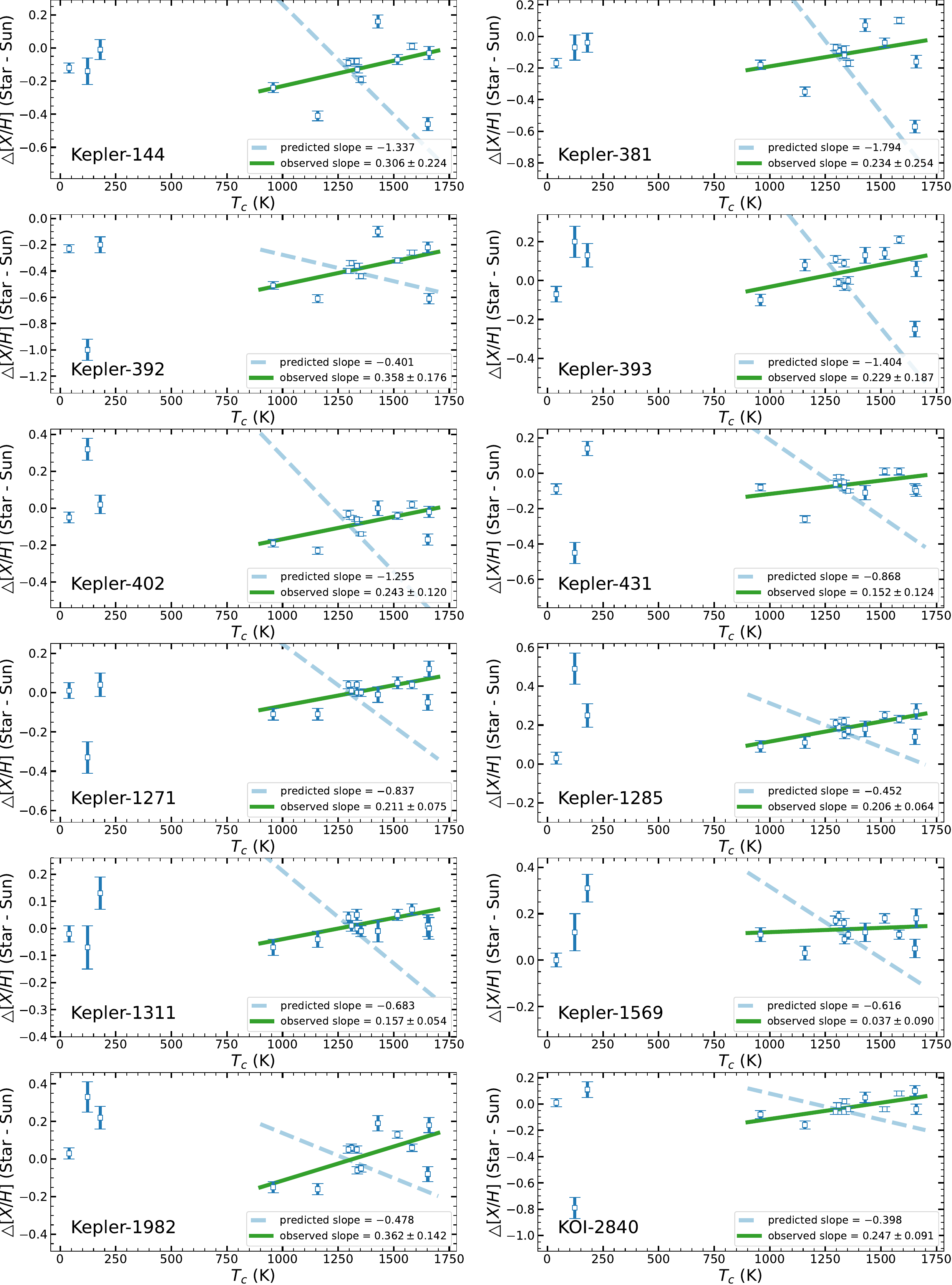}
\caption{Photospheric elemental abundances from the combined
\citet{2016ApJS..225...32B}/\citet{2018ApJS..237...38B} catalog
as a function of $T_{\text{c}}$ for the 10 Kepler-discovered
terrestrial exoplanet systems that we predict to show the strongest
refractory depletion signals.  We plot as the dashed blue lines
the \citet{2010ApJ...724...92C}-predicted relationships.  For the
refractory elements with $T_{\text{c}} > 900$ K, we fit using
weighted least squares linear relationships between $T_{\text{c}}$ and
photospheric elemental abundance and plot those results as the solid
green lines.  None of the observed lines are statistically consistent
with the \citet{2010ApJ...724...92C}-predicted slopes given the total
observed terrestrial exoplanet mass in each system.  We conclude that
terrestrial planet formation does not in general result in the depletion
of refractory elements in the photospheres of stars hosting terrestrial
planet systems.\label{fig:slope}}
\end{figure*}

\begin{figure*}
\plotone{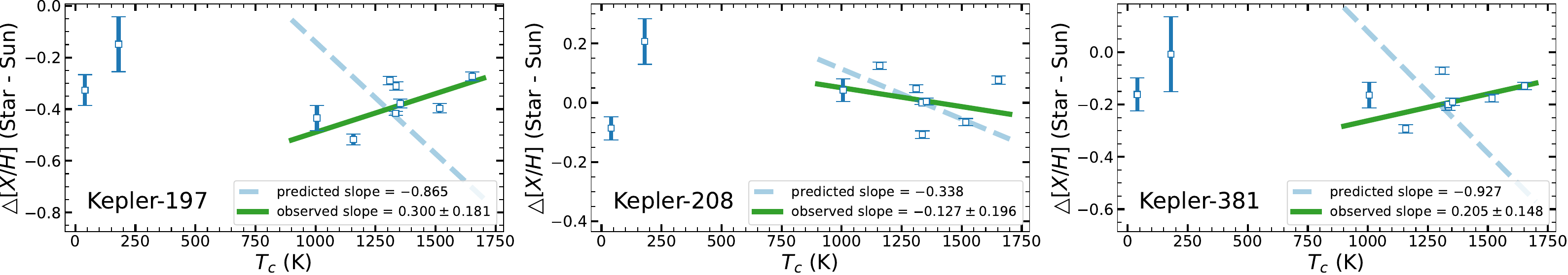}
\caption{Photospheric elemental abundances from APOGEE DR17
as a function of $T_{\text{c}}$ for the four Kepler-discovered
terrestrial exoplanet systems that we predict to show the strongest
refractory depletion signals.  We plot as the dashed blue lines
the \citet{2010ApJ...724...92C}-predicted relationships.  For the
refractory elements with $T_{\text{c}} > 900$ K, we fit using
weighted least squares linear relationships between $T_{\text{c}}$ and
photospheric elemental abundance and plot those results as the solid
green lines.  None of the observed lines are statistically consistent
with the \citet{2010ApJ...724...92C}-predicted slopes given the total
observed terrestrial exoplanet mass in each system.  This observation
validates our Brewer-based inference and confirms the robustness of our
conclusion.\label{fig:apogee}}
\end{figure*}

The elemental abundances inferences available in the combined
\citet{2016ApJS..225...32B}/\citet{2018ApJS..237...38B}
or APOGEE DR17 catalogs have precisions of about 0.03 dex
and are therefore slightly less precise than those inferred
deferentially for solar twins with precisions of about 0.01 dex
\citep[e.g.,][]{2009ApJ...704L..66M,2018ApJ...865...68B}.  We therefore
focus our attention on the terrestrial exoplanet system host stars
expected to have the largest signals of refractory depletion easily
discernible even with elemental abundance inference precisions of
0.03 dex.  We select systems with $S_{\text{pred}} \leq -0.25$ and focus
on 13 terrestrial exoplanet systems with elemental abundances from the
combined \citet{2016ApJS..225...32B}/\citet{2018ApJS..237...38B} catalog
and three from the APOGEE DR17 catalog.  We calculate $S_{\text{obs}}$
and its uncertainty for each of these systems by weighted least-squares
fitting a linear function relating elements' condensation temperatures
to the inferred elemental abundances in the photosphere of each system's
host star for elements with $T_{\text{c}} > 900$ K.  We illustrate the
results of these calculations in Table \ref{table:metalPoor} and Figures
\ref{fig:Kepler}, \ref{fig:slope}, and \ref{fig:apogee}.

\begin{deluxetable*}{ccccccc}
\tablecaption{Kepler-discovered Terrestrial Planet Systems with Elemental
Abundances Expected to Show the Strongest Refractory Depletion Signals}
\label{table:metalPoor}
\tablehead{
\colhead{Simbad ID} &
\colhead{$M_{\text{cz}}$} &
\colhead{[M/H]} &
\colhead{Terrestrial Planet Mass} &
\colhead{$S_{\text{obs}}$} &
\colhead{$S_{\text{pred}}$} &
\colhead{$p$-value} \\
\colhead{} &
\colhead{($M_{\odot}$)} &
\colhead{} &
\colhead{($M_{\oplus}$)} &
\colhead{(dex K$^{-1}$)} &
\colhead{(dex K$^{-1}$)} &
\colhead{}
}
\startdata
\multicolumn{7}{l}{\textbf{Combined \citet{2016ApJS..225...32B}/\citet{2018ApJS..237...38B} catalog}}\\
Kepler-144  & 0.0015 & $-0.12$ &  6.0 & $+0.306\pm0.224$ & $-1.337$ & $1.27\times10^{-5}$ \\
Kepler-215  & 0.0145 & $-0.48$ &  7.6 & $+0.187\pm0.135$ & $-0.263$ & $3.90\times10^{-3}$ \\
Kepler-381  & 0.0004 & $-0.10$ &  3.1 & $+0.234\pm0.254$ & $-1.794$ & $5.99\times10^{-6}$ \\
Kepler-392  & 0.0039 & $-0.37$ &  3.0 & $+0.358\pm0.176$ & $-0.401$ & $7.56\times10^{-4}$ \\
Kepler-393  & 0.0013 & $+0.03$ &  7.3 & $+0.229\pm0.187$ & $-1.404$ & $2.68\times10^{-6}$ \\
Kepler-402  & 0.0027 & $-0.06$ & 10.7 & $+0.243\pm0.120$ & $-1.255$ & $1.04\times10^{-7}$ \\
Kepler-431  & 0.0021 & $-0.06$ &  5.2 & $+0.152\pm0.124$ & $-0.868$ & $4.74\times10^{-6}$ \\
Kepler-1271 & 0.0030 & $+0.02$ &  7.9 & $+0.211\pm0.075$ & $-0.837$ & $3.43\times10^{-8}$ \\
Kepler-1285 & 0.0013 & $+0.18$ &  2.2 & $+0.206\pm0.064$ & $-0.452$ & $6.33\times10^{-7}$ \\
Kepler-1311 & 0.0045 & $+0.03$ &  9.5 & $+0.157\pm0.054$ & $-0.683$ & $1.28\times10^{-8}$ \\
Kepler-1569 & 0.0039 & $+0.13$ &  8.3 & $+0.037\pm0.090$ & $-0.616$ & $1.37\times10^{-5}$ \\
Kepler-1982 & 0.0010 & $+0.00$ &  1.5 & $+0.362\pm0.142$ & $-0.478$ & $7.49\times10^{-5}$ \\
KOI-2840    & 0.0092 & $-0.02$ & 10.8 & $+0.247\pm0.091$ & $-0.398$ & $1.62\times10^{-5}$ \\
\hline
\multicolumn{7}{l}{\textbf{APOGEE DR17 catalog}}\\
Kepler-197  & 0.0071 & $-0.42$ & 10.6 & $+0.300\pm0.181$ & $-0.865$ & $3.32\times10^{-4}$ \\
Kepler-208  & 0.0090 & $-0.00$ &  9.9 & $-0.127\pm0.196$ & $-0.338$ & $1.62\times10^{-1}$ \\
Kepler-381  & 0.0014 & $-0.20$ &  3.0 & $+0.205\pm0.148$ & $-0.927$ & $1.31\times10^{-4}$ \\
\enddata
\end{deluxetable*}

Contrary to the suggestion from \citet{2009A&A...508L..17R}, as we
present in Table \ref{table:metalPoor} and Figures \ref{fig:Kepler},
\ref{fig:slope}, and \ref{fig:apogee} in general we find negative
slopes for the linear relation between between photospheric elemental
abundance and $T_{\text{c}}$ for the individual host stars of
these terrestrial exoplanet systems.  Indeed, when we compare our
inferred $S_{\text{obs}}$ values with the $S_{\text{pred}}$ values
from the \citet{2010ApJ...724...92C} formalism using one-sample
$t$-tests\footnote{With 10 degrees of freedom for the combined
\citet{2016ApJS..225...32B}/\citet{2018ApJS..237...38B} sample and six
from the APOGEE DR17 sample (i.e., 12/8 elemental abundances minus
two model parameters)}, we consistently reject the null hypothesis
that $S_{\text{obs}} = S_{\text{pred}}$ with very small $p$-values
(Table \ref{table:metalPoor}).  To evaluate the significance of this
observation over our entire sample, we use the formula
\begin{equation}
P = n!\times\Pi_{i=1}^{n} p_{i},
\label{eqn:prob}
\end{equation}
where $P$ is the probability of the null hypothesis $S_{\text{obs}}
= S_{\text{pred}}$ marginalized over our entire sample and
$n$ is the number of systems in our sample and $p_{i}$ are
the individual $p$-values.  Aggregating over our individual
\citet{2016ApJS..225...32B}/\citet{2018ApJS..237...38B} and APOGEE
DR17 samples as well their union, we find $p = (8.52 \times 10^{-60},
4.22 \times 10^{-8}, 2.02 \times 10^{-64})$.  In words, even for the
systems expected to have the largest signals, the probability that
the observed photospheric elemental abundances of Kepler-observed
terrestrial exoplanet system host stars matches the prediction from the
\citet{2009A&A...508L..17R} refractory sequestration scenario as modeled
by \citet{2010ApJ...724...92C} is essentially zero.

\subsection{Gap-opening Giant Exoplanet Formation}\label{sec:gapopening}

To evaluate the possible impact of the formation of gap-opening
giant exoplanets on the elemental abundances of their host stars,
we use the sample of 60 gap-opening giant exoplanet stars and
224 single stars with elemental abundances from the combined
\citet{2016ApJS..225...32B}/\citet{2018ApJS..237...38B} catalog
uniformly searched for giant exoplanets with the Doppler technique by
\citet{2021ApJS..255....8R} described in Section \ref{sec:sample}.  As in
Section \ref{sec:tpsystem}, we focus on the subset of giant exoplanet
host stars with the least-massive outer convective zones expected to
display the strongest signals of planet-modulated host star photospheric
elemental abundances.

To that end, we use the $M_{\text{cz}}$ evolution histories we generated
with MESA in Section \ref{sec:Surfcz} and interpolate $M_{\text{cz}}$
as a function of stellar mass and metallicity at 2 Myr, a typical disk
lifetime for short-lived disks with high X-ray photoevaporation rates
\citep[e.g.,][]{2020MNRAS.493.5079B}.  We focus on the 28 systems that
have host star $M_{\text{cz}} < 0.8~M_{\ast}$ at 2 Myr.  If the formation
of Jupiter and the gap it opened in the protosolar nebula was responsible
for the solar photosphere's low refractory-to-volatile ratio as suggested
by \citet{2020MNRAS.493.5079B} and \citet{2023A&A...676A..87H}, then
the host stars of these systems should have the most readily observable
depleted refractory-to-volatile ratios among the known giant systems that,
like the solar system, experienced gap opening by a giant planet.

To isolate the impact of the formation of gap-opening giant exoplanets
from possible confounders like stellar or Galactic chemical evolution,
for each giant exoplanet system we first select from the sample of
single stars that has been uniformly searched for giants exoplanets
the closest ``analog'' with the smallest Euclidean distance in stellar
mass/metallicity space from the subsamples of stars with stellar
metallicities within 0.1 dex and stellar masses with $0.05~M_{\odot}$.
We provide these 28 giant exoplanet systems and their analogs in Table
\ref{table:giant}.  Using the photospheric elemental abundances form
the combined \citet{2016ApJS..225...32B}/\citet{2018ApJS..237...38B}
catalog, we use a Monte Carlo simulation to quantify the degree of
refractory depletion in this sample of $n = 28$ giant exoplanet systems.
For each giant exoplanet host and its analog, we first calculate mean
photospheric abundances of volatile and refractory elements
\begin{eqnarray}
\bar{x}_{\text{vol}}    & = & \frac{1}{3} \sum_{i=1}^{3} [\text{X}_{i}/\text{H}], \\
\bar{x}_{\text{refrac}} & = & \frac{1}{12} \sum_{j=1}^{12} [\text{X}_{j}/\text{H}],
\end{eqnarray}
where $X_{i} \in (\text{C,N,O})$ and\\$X_{j} \in
(\text{Na,Mg,Al,Si,Ca,Ti,V,Cr,Mn,Fe,Ni,Y})$.  For each star, we then
calculate the degree of refractory depletion as the difference $\Delta =
\bar{x}_{\text{vol}}-\bar{x}_{\text{refrac}}$ between those two values.
For each pair, we next calculate $\Delta^{2} = \Delta_{\text{host}} -
\Delta_{\text{analog}}$ and what we term the relative refractory offset
parameter $R = (1/n) \sum_{i=1}^{n} \Delta^2$.  We repeat this process
200 times using the elemental abundance uncertainties in the combined
\citet{2016ApJS..225...32B}/\citet{2018ApJS..237...38B} catalog to
infer the distributions of $\Delta$ for each giant exoplanet host and
its analog as well as the distribution of $R$ for the entire sample.

\begin{deluxetable*}{cccccccc}[t]
\caption{Giant Planet Hosting Systems and Their Closest Analogs \label{table:giant}}
\tablehead{\colhead{Simbad ID} &
\colhead{Star Mass} &
\colhead{[M/H]} &
\colhead{$M_{\text{cz}}/M_{\ast}$} &
\colhead{Closest Analog} &
\colhead{Analog} &
\colhead{Analog} &
\colhead{Analog $M_{\text{cz}}/M_{\ast}$}\\
\colhead{} &
\colhead{$(M_{\odot})$} &
\colhead{} &
\colhead{at 2 Myr} &
\colhead{Simbad ID} &
\colhead{Mass} &
\colhead{[M/H]} &
\colhead{at 2 Myr} \\
\colhead{} &
\colhead{} &
\colhead{} &
\colhead{} &
\colhead{} &
\colhead{$(M_{\odot})$} &
\colhead{} &
\colhead{}}
\decimals
\startdata
$\ast$ 109 Psc & 1.13 & $+0.11$ & 0.776 & $\ast$ 79 Cet & 1.12 & $+0.11$ & 0.781 \\
$\ast$ 70 Vir & 1.09 & $-0.07$ & 0.747 & HD 157338 & 1.06 & $-0.08$ & 0.759 \\
HD 11964 & 1.10 & $+0.09$ & 0.783 & HD 13043 & 1.14 & $+0.09$ & 0.766 \\
HD 136925 & 0.96 & $-0.20$ & 0.786 & HD 20619 & 0.95 & $-0.20$ & 0.792 \\
HD 13931 & 1.08 & $+0.06$ & 0.783 & HD 142229 & 1.07 & $+0.06$ & 0.788 \\
$\ast$ rho CrB & 1.00 & $-0.18$ & 0.771 & $\ast$ bet CVn & 0.99 & $-0.18$ & 0.775 \\
HD 169830 & 1.41 & $+0.17$ & 0.663 & HD 63754 & 1.41 & $+0.21$ & 0.672 \\
HD 170469 & 1.14 & $+0.28$ & 0.800 & HD 144585 & 1.17 & $+0.29$ & 0.790 \\
HD 178911B & 1.34 & $+0.20$ & 0.701 & HD 21847 & 1.30 & $+0.23$ & 0.726 \\
HD 179949 & 1.23 & $+0.20$ & 0.749 & HD 9562 & 1.26 & $+0.21$ & 0.739 \\
HD 183263 & 1.17 & $+0.27$ & 0.788 & $\ast$ mu.01 Her & 1.13 & $+0.26$ & 0.800 \\
HD 187123 & 1.08 & $+0.09$ & 0.792 & HD 110537 & 1.04 & $+0.09$ & 0.809 \\
HD 195019 & 1.09 & $+0.04$ & 0.774 & $\ast$ 88 Leo & 1.11 & $+0.04$ & 0.765 \\
HD 209458 & 1.15 & $+0.05$ & 0.749 & HD 68978 & 1.10 & $+0.06$ & 0.774 \\
HD 222582 & 1.04 & $+0.01$ & 0.790 & HD 129814 & 1.06 & $+0.01$ & 0.781 \\
HD 24040 & 1.12 & $+0.17$ & 0.790 & HD 26736 & 1.10 & $+0.17$ & 0.799 \\
HD 33636 & 1.06 & $-0.07$ & 0.762 & HD 157338 & 1.06 & $-0.08$ & 0.759 \\
HD 34445 & 1.14 & $+0.11$ & 0.773 & $\ast$ 79 Cet & 1.12 & $+0.11$ & 0.781 \\
HD 37124 & 0.86 & $-0.35$ & 0.796 & HD 68017 & 0.86 & $-0.35$ & 0.796 \\
HD 40979 & 1.23 & $+0.24$ & 0.756 & HD 21313 & 1.23 & $+0.23$ & 0.755 \\
HD 4208 & 0.90 & $-0.26$ & 0.798 & $\ast$ nu.02 Lup & 0.92 & $-0.26$ & 0.787 \\
HD 50499 & 1.24 & $+0.31$ & 0.765 & HD 6558 & 1.24 & $+0.28$ & 0.760 \\
HD 50554 & 1.11 & $+0.02$ & 0.759 & HD 1388 & 1.09 & $+0.02$ & 0.768 \\
HD 52265 & 1.21 & $+0.20$ & 0.758 & HD 21313 & 1.23 & $+0.23$ & 0.755 \\
HD 74156 & 1.25 & $+0.11$ & 0.723 & HD 114613 & 1.27 & $+0.15$ & 0.722 \\
HD 82943 & 1.17 & $+0.26$ & 0.784 & $\ast$ mu.01 Her & 1.13 & $+0.26$ & 0.800 \\
HD 8574 & 1.17 & $+0.03$ & 0.735 & HD 50639 & 1.16 & $+0.03$ & 0.739 \\
$\ast$ 47 UMa & 1.09 & $+0.04$ & 0.774 & $\ast$ 88 Leo & 1.11 & $+0.04$ & 0.765
\enddata
\end{deluxetable*}

\begin{figure*}
\plotone{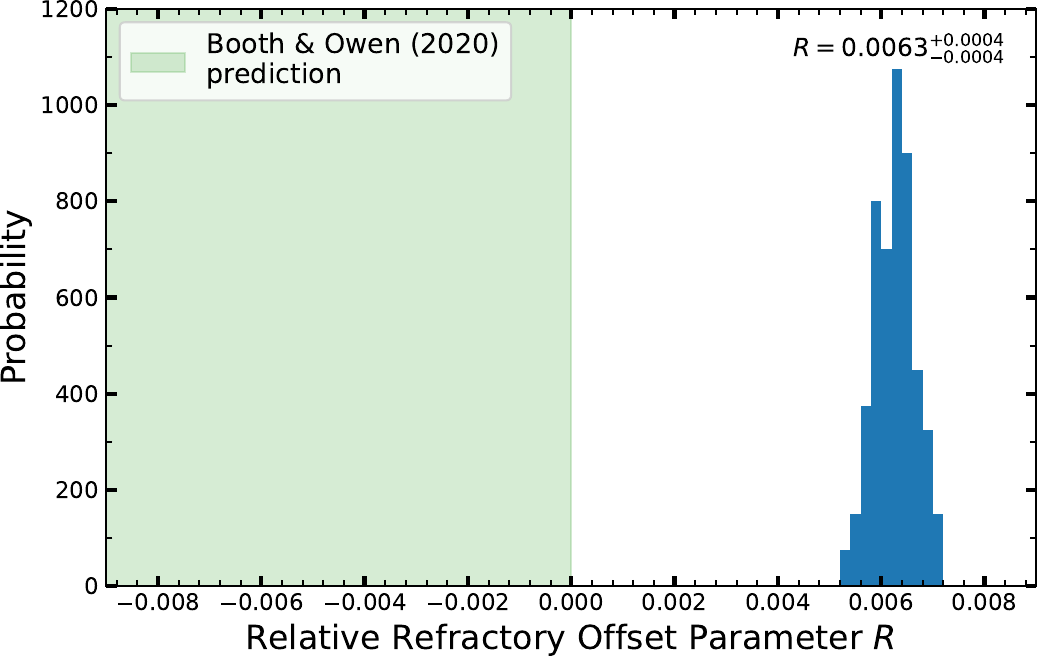}
\caption{Histogram showing the probability density function for the
$R$ parameter we define as the difference between the refractory
depletion of a giant exoplanet host star and its analogous star
without a detected giant exoplanet.  While the distribution excludes
zero, it is shifted to positive values.   This suggests that giant
exoplanet host stars are more refractory rich than their analogous
stars without detected giant exoplanets.  The enhanced refractory
abundances of giant exoplanet host stars is inconsistent with the
\citet{2020MNRAS.493.5079B}/\citet{2023A&A...676A..87H} scenario in
which the formation of giant planets is responsible for the apparent
differences between the refractory abundances of the Sun and nearby
solar analogs.\label{fig:giant}}
\end{figure*}

If the formation of a giant planet like Jupiter capable of opening a gap
in its parent protoplanetary disk prevents the accretion of refractory
material onto its host star as suggested by \citet{2020MNRAS.493.5079B}
and \citet{2023A&A...676A..87H}, then $R$ should be negative.  As we plot
in Figure \ref{fig:giant}, this is not the case.  As a result, the solar
photosphere's low refractory-to-volatile ratio cannot be explain by the
formation of Jupiter and the gap it opened in the protosolar nebula as
suggested by \citet{2020MNRAS.493.5079B} and \citet{2023A&A...676A..87H}.
Indeed, it seems that giant exoplanet host stars are actually enhanced
in the abundances of the refractory elements by about $0.0063\pm 0.0004$
dex, possibly a higher-order consequence of the well known connection
between host star total metallicity and giant exoplanet occurrence.

\begin{figure*}
\plotone{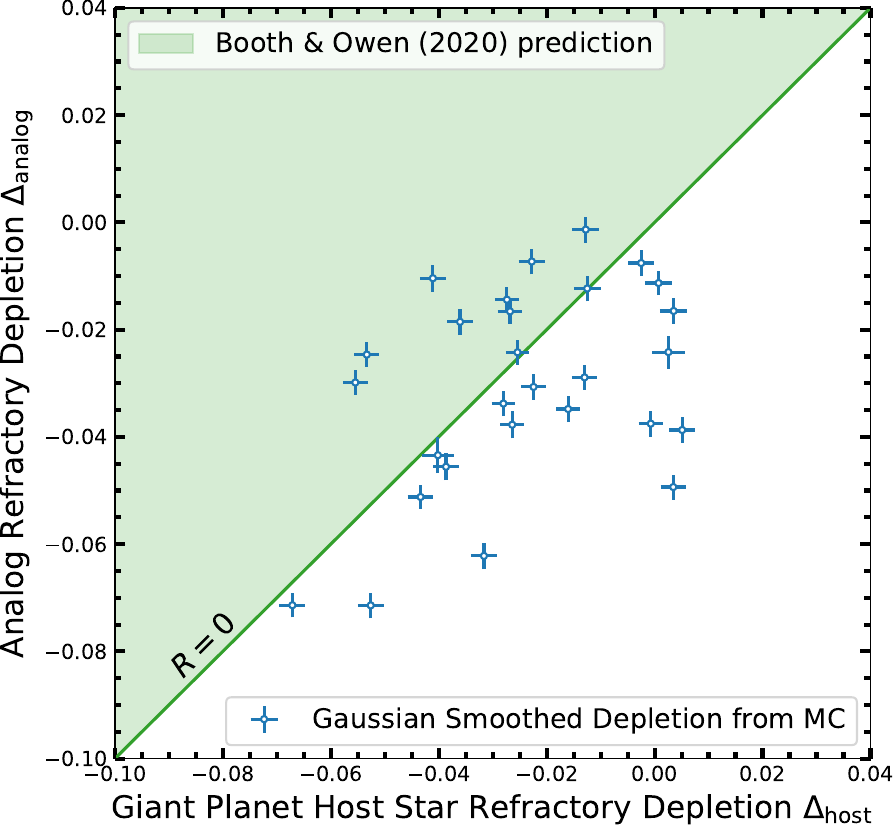}
\caption{Distribution of Gaussian-smoothed refractory depletions
for both giant exoplanet host stars and their closest analog
single stars based on 200-iteration Monte Carlo (MC) simulations.
Each point represents the median of a Monte Carlo simulation for an
individual (giant exoplanet host star, closest analogous star without
detected giant exoplanet) pair.  The error bars correspond to a Monte
Carlo simulation's quantiles corresponding to its 3-$\sigma$ range.
The \citet{2020MNRAS.493.5079B}/\citet{2023A&A...676A..87H} scenario
predicts that most points should be in the shaded region above and to
left of the line defined by the relative refractory depletion offset $R
= 0$.  These points instead preferentially lie below and to the right
of the $R = 0$ line, suggesting that giant exoplanet host stars are a
refractory-rich population.\label{fig:giant_scatter}}
\end{figure*}

We more closely examine the outcome the Monte Carlo simulation described
above by applying a two-dimensional Gaussian filter to the distributions
of $\Delta_{\text{host}}$ and $\Delta_{\text{analog}}$.  If points in
the $\Delta_{\text{host}}$--$\Delta_{\text{analog}}$ plane lie above
and to the left of the line $y=x$, then the formation of a giant planet
capable of opening a gap in its parent protoplanetary disk is associated
with low refractory abundances.  In accord with our $R$-based conclusion
in Figure \ref{fig:giant}, as we plot in Figure \ref{fig:giant_scatter}
the majority of stars in our sample lie below and to the right of the
line $y=x$ in the $\Delta_{\text{host}}$--$\Delta_{\text{analog}}$ plane.
This observation reinforces the conclusion that the host stars of giant
exoplanets capable of opening a gap in their parent protoplanetary disks
are not refractory depleted but rather slightly refractory enriched.

\section{Discussion}\label{sec:discussion}

As we showed in Section \ref{sec:Surfcz}, solar-mass, solar-metallicity
stars have massive surface convective zones for the entire durations
of any protoplanetary disks they once hosted.  As we showed in Figure
\ref{fig:evolution}, our solar model ceases to be fully convective after
about 1 Myr and from that point its surface convective zone shrinks
over 30 Myr to the approximately constant mass it will have for the
main sequence phase of that solar model's evolution.  During the era of
protoplanetary disk in the first 10 Myr of our solar model's evolution
when plantesimals, oligarchs, and giant planets form, the solar model
always has $M_{\text{cz}} \geq 0.35~M_{\odot}$.  In this case, the
sequestration of more than $50~M_{\oplus}$ of rocky material would be
needed to reproduce the solar photosphere's low refractory-to-volatile
ratio.  This mass is well in excess of the total amount of rocky material
inside the water-ice line of the minimum-mass solar nebula (about
$4~M_{\oplus}$), and this analysis by itself indicates that terrestrial
planet formation during the era of protoplanetary disks cannot be the
explanation for abundance anomalies in the photospheres of solar-mass,
solar-metallicity stars.

We neglected the impact of disk accretion on stellar structure in
our simulations.  While this a good approximation for low-mass disks
during the era of planet formation with consistent mass accretion rates
$\dot{M}$ in the range $10^{-9} M_{\odot}~\text{yr}^{-1} \lesssim
\dot{M} \lesssim 10^{-7} M_{\odot}~\text{yr}^{-1}$, higher mass
accretion rates and or time variability could change stellar structure
\citep[e.g.,][]{2010A&A...521A..44B,2017A&A...599A..49K}.  Nevertheless,
\citet{2018A&A...618A.132K} showed that more comprehensive modeling yields
the same surface convective zones masses to within a factor of two of
those predicted by simpler models like those we use in this analysis.

Using two independent catalogs of high-quality photospheric
elemental abundances, we find that the host stars of
Kepler-discovered terrestrial exoplanet systems do not share
the solar photosphere's inverse relationship between elemental
abundance and condensation temperature that has often been
attributed to the formation of the solar system's terrestrial planets
\citep[e.g.,][]{2009ApJ...704L..66M,2009A&A...508L..17R}.  This is true
even among the host stars with the lowest-mass outer convective zones
expected to show the most easily observed evidence of this property of
the solar photosphere.  This suggests that terrestrial planet formation
and thereby the sequestration of significant fractions of protoplanetary
disks' initial complements of rocky material in terrestrial exoplanets
does not lead to inverse relationships between elemental abundance
and condensation temperature in the photospheres of their host stars.
Other explanations like Galactic chemical evolution must be the reason
for the differences between the photospheric elemental abundances of
the Sun and nearby solar twins.  \citet{2021MNRAS.507.2220G} reached
a similar conclusion for Kepler-discovered exoplanets of all radii
using a smaller sample of stars with photospheric elemental abundances
from \citet{2016ApJS..225...32B} and \citet{2018ApJS..237...38B}
as did \citet{2026ApJ...998..301G} based on their own analyses of the
Keck/HIRES spectra for Kepler-discovered exoplanet host stars underlying
the \citet{2016ApJS..225...32B} and \citet{2018ApJS..237...38B} catalogs.
Because they included exoplanets of all radii, they were unable to isolate
the effect of terrestrial exoplanet formation as we have in this study.

In a sample of stars uniformly searched for giant exoplanets, brown
dwarfs, and binary stars using the Doppler technique, we find that
the host stars of giant exoplanets capable of opening a gap in their
parent protoplanetary disks do not share the solar photosphere's low
refractory-to-volatile ratio that has been associated with the formation
of such planets \citep[e.g.,][]{2020MNRAS.493.5079B,2023A&A...676A..87H}.
This is true even among the host stars with the lowest-mass surface
convective zones expected to show the most easily observed evidence of
refractory depletion.  In contrast, we find evidence that the photospheres
of the host stars of giant exoplanets capable of opening a gap in their
parent protoplanetary disks are slightly enriched in refractory elements.
This observation implies that the formation of giant planets capable
of opening gaps in their parent protoplanetary disks and thereby the
sequestration of large amounts of rocky material outside those gaps
does not lead to low refractory-to-volatile ratios in the photospheres
of their host stars.

Our results reflect the now long-known increase in the occurrence of giant
planets orbiting refractory-rich host stars at fixed iron metallicity
\cite[e.g.,][]{2011ApJ...738...97B,2012A&A...543A..89A}.  This increase in
the occurrence of giant planets orbiting refractory-rich stars at constant
metallicity has been explained in the same way as the even more well-known
relationship between total host star metallicity and giant planet
occurrence \citep[e.g.,][]{2004A&A...415.1153S,2005ApJ...622.1102F}:
more planet-making material leads to more giant planets.

\section{Conclusion}\label{sec:conclusion}

To evaluate the possible impact of the formation of systems of terrestrial
planets on the photospheric elemental abundances of their host stars,
we isolate the Kepler-discovered terrestrial exoplanet system host
stars with the lowest-mass surface convective zones (both now and
during the era of planet formation) expected to show the most easily
detectable depletions of refractory elements and/or inverse relationships
between photospheric elemental abundance and condensation temperature.
If the formation of systems of terrestrial exoplanets sequester rocky
material that would otherwise accrete onto the surfaces of their host
stars, then these host stars should share the solar photosphere's
low refractory-to-volatile ratio and its inverse relationship between
elemental abundance and condensation temperature that has often been
attributed to the formation of the solar system's terrestrial planets.
To mitigate possible systematics, we perform our analyses on two
independent catalogs of high-quality elemental abundance inferences for
these terrestrial exoplanet host stars.  We ultimately find no evidence
that Kepler-discovered terrestrial exoplanet system host stars evince
inverse relationships between photospheric elemental abundance and
condensation temperature.  Some other process unrelated to terrestrial
planet formation must be responsible for the solar photosphere's
apparently low refractory-to-volatile ratio and its inverse relationship
between elemental abundance and condensation temperature.

Similarly, to assess the possible impact of the formation of giant
planets capable of opening a gap in their parent protoplanetary disk on
the photospheric elemental abundances of their host stars, we define
a sample of stars uniformly searched with the Doppler technique for
giant exoplanets, brown dwarfs, and binary stars with the least-massive
surface convective zones expected to show the most easily discernible
depletions of refractory elements.  If the gap in a protoplanetary
disk created by a newly formed giant planet interdicts the inward
migration of rocky material, then $\sim\!\!100~M_{\oplus}$ of rocky
material could be sequestered in the outer disk and thereby prevented
from accreting onto the host of the newly formed giant planet.  We find
no evidence that the host stars of giant exoplanets capable of opening
gaps in their parent protoplanetary disks when compared to stars with
similar masses, metallicities, and ages.  Instead, we find that the
photospheres of the host stars of giant exoplanets capable of opening
gaps are enhanced in abundances of the refractory elements by about
$0.0063\pm0.0004$ dex on average.  We conclude that the sequestration
of refractory elements in terrestrial exoplanet systems or in outer
disks as a consequence of giant planets does not systemically affect
the photospheric refractory-to-volatile ratios of their host stars.

\begin{acknowledgments}

Le-Chris Wang was supported by Johns Hopkins University through a
Summer Provost's Undergraduate Research Award and a Dean's Arts and
Sciences Projects, Investigations, and Research Endeavors (ASPIRE) grant.
This material is based upon work supported by the National Aeronautics
and Space Administration under Grant No. 80NSSC23K0266 issued through the
Exoplanets Research Program (XRP).  Funding for SDSS-III has been provided
by the Alfred P. Sloan Foundation, the Participating Institutions, the
National Science Foundation, and the U.S. Department of Energy Office
of Science.  The SDSS-III web site is \url{http://www.sdss3.org/}.
SDSS-III is managed by the Astrophysical Research Consortium for the
Participating Institutions of the SDSS-III Collaboration including the
University of Arizona, the Brazilian Participation Group, Brookhaven
National Laboratory, Carnegie Mellon University, University of Florida,
the French Participation Group, the German Participation Group, Harvard
University, the Instituto de Astrofisica de Canarias, the Michigan
State/Notre Dame/JINA Participation Group, Johns Hopkins University,
Lawrence Berkeley National Laboratory, Max Planck Institute for
Astrophysics, Max Planck Institute for Extraterrestrial Physics, New
Mexico State University, New York University, Ohio State University,
Pennsylvania State University, University of Portsmouth, Princeton
University, the Spanish Participation Group, University of Tokyo,
University of Utah, Vanderbilt University, University of Virginia,
University of Washington, and Yale University.  Funding for the Sloan
Digital Sky Survey IV has been provided by the Alfred P. Sloan Foundation,
the U.S. Department of Energy Office of Science, and the Participating
Institutions.  SDSS-IV acknowledges support and resources from the Center
for High Performance Computing  at the University of Utah.  Funding for
the Sloan Digital Sky Survey IV has been provided by the Alfred P. Sloan
Foundation, the U.S. Department of Energy Office of Science, and the
Participating Institutions.  SDSS-IV acknowledges support and resources
from the Center for High Performance Computing  at the University of Utah.
The SDSS website is \url{www.sdss4.org}.  SDSS-IV is managed by the
Astrophysical Research Consortium for the Participating Institutions of
the SDSS Collaboration including the Brazilian Participation Group, the
Carnegie Institution for Science, Carnegie Mellon University, Center for
Astrophysics | Harvard \& Smithsonian, the Chilean Participation Group,
the French Participation Group, Instituto de Astrof\'isica de Canarias,
The Johns Hopkins University, Kavli Institute for the Physics and
Mathematics of the Universe (IPMU) / University of Tokyo, the Korean
Participation Group, Lawrence Berkeley National Laboratory, Leibniz
Institut f\"ur Astrophysik Potsdam (AIP), Max-Planck-Institut f\"ur
Astronomie (MPIA Heidelberg), Max-Planck-Institut f\"ur Astrophysik
(MPA Garching), Max-Planck-Institut f\"ur Extraterrestrische Physik
(MPE), National Astronomical Observatories of China, New Mexico State
University, New York University, University of Notre Dame, Observat\'ario
Nacional / MCTI, The Ohio State University, Pennsylvania State University,
Shanghai Astronomical Observatory, United Kingdom Participation Group,
Universidad Nacional Aut\'onoma de M\'exico, University of Arizona,
University of Colorado Boulder, University of Oxford, University of
Portsmouth, University of Utah, University of Virginia, University of
Washington, University of Wisconsin, Vanderbilt University, and Yale
University.  This work has made use of data from the European Space
Agency (ESA) mission Gaia (\url{https://www.cosmos.esa.int/gaia}),
processed by the Gaia Data Processing and Analysis Consortium (DPAC,
\url{https://www.cosmos.esa.int/web/gaia/dpac/consortium}).  Funding for
the DPAC has been provided by national institutions, in particular
the institutions participating in the Gaia Multilateral Agreement.
This research has made use of the NASA Exoplanet Archive, which is
operated by the California Institute of Technology, under contract with
the National Aeronautics and Space Administration under the Exoplanet
Exploration Program.  This research has made use of the SIMBAD database,
operated at CDS, Strasbourg, France \citep{2000A&AS..143....9W}.
This research has made use of the VizieR catalog access tool, CDS,
Strasbourg, France.  The original description of the VizieR service
was published in \citet{2000A&AS..143...23O}.  This research has made
use of the Astrophysics Data System, funded by NASA under Cooperative
Agreement 80NSSC21M0056.

\end{acknowledgments}

\facilities{Exoplanet Archive, Gaia, Keck:I (HIRES), Kepler, Sloan}

\software{\texttt{astropy} \citep{2013A&A...558A..33A, 2018AJ....156..123A},
          \texttt{matplotlib} \citep{4160265},
          \texttt{numpy} \citep{2020Natur.585..357H},
          \texttt{pandas} \citep{mckinney-proc-scipy-2010, reback2020pandas},
          \texttt{R} \citep{r26},
          \texttt{scipy} \citep{2020SciPy-NMeth},
          \texttt{TOPCAT} \citep{2005ASPC..347...29T}}

\section*{Data Availability}  

The MESA inlist and related output files for each star
modeled in this work are also publicly available on Zenodo:
\dataset[10.5281/zenodo.23254381]{https://doi.org/10.5281/zenodo.23254381}.

\bibliography{article02}{}
\bibliographystyle{aasjournalv7}

\end{CJK*}
\end{document}